\documentclass[useAMS,referee,usenatbib,usegraphicx]{biom}

\usepackage{times}

\def\bSig\mathbf{\Sigma}

\usepackage{lineno}
\usepackage[none]{hyphenat}

\usepackage{graphicx}
\DeclareGraphicsExtensions{.png,.pdf}
\graphicspath{{./Figures/}}

\usepackage{color}
\usepackage{verbatim}

\usepackage{booktabs}
\usepackage{float}
\usepackage{placeins}
\usepackage{enumitem}
\usepackage{bm, amsmath, amssymb, url}
\usepackage[titletoc]{appendix}

\title[]{Efficient Bayesian hierarchical functional data analysis with basis function approximations using Gaussian-Wishart processes}

\author{Jingjing Yang \emailx{yjingj@umich.edu} \\
Department of Biostatistics, University of Michigan, Ann Arbor, MI 48109, USA.
\and Dennis D.~Cox\\
Department of Statistics, Rice University, Houston, TX 77005, USA.
\and Jong Soo Lee \\
Department of Mathematical Sciences, University of Massachusetts Lowell, Lowell, MA 01854, USA.
\and  Peng Ren \\
Suntrust Banks Inc, Atlanta, GA 30308, USA.
\and Taeryon Choi \emailx{trchoi@korea.ac.kr} \\
Department of Statistics, Korea University, Seoul 136-701, Republic of Korea.
}

\begin{document}
{\sloppy
\label{firstpage}

\begin{abstract}
Functional data are defined as realizations of random functions (mostly smooth functions) varying over a continuum, which are usually collected with measurement errors on discretized grids. In order to accurately smooth noisy functional observations and deal with the issue of high-dimensional observation grids, we propose a novel Bayesian method based on the Bayesian hierarchical model with a Gaussian-Wishart process prior and basis function representations. We first derive an induced model for the basis-function coefficients of the functional data, and then use this model to conduct posterior inference through Markov chain Monte Carlo. Compared to the standard Bayesian inference that suffers serious computational burden and unstableness for analyzing high-dimensional functional data, our method greatly improves the computational scalability and stability, while inheriting the advantage of simultaneously smoothing raw observations and estimating the mean-covariance functions in a nonparametric way. In addition, our method can naturally handle functional data observed on random or uncommon grids. Simulation and real studies demonstrate that our method produces similar results as the standard Bayesian inference with low-dimensional common grids, while efficiently smoothing and estimating functional data with random and high-dimensional observation grids where the standard Bayesian inference fails. In conclusion, our method can efficiently smooth and estimate high-dimensional functional data, providing one way to resolve the curse of dimensionality for Bayesian functional data analysis with Gaussian-Wishart processes.
\end{abstract}


\maketitle


\section{Introduction}
\label{intro}
Functional data --- defined as realizations of random functions varying over a continuum \citep{ramsay2005functional} --- include a variety of data types such as longitudinal data, spatial-temporal data, and image data. Because functional data are generally collected on discretized grids with measurement errors, constructing functions from noisy discrete observations (referred to as smoothing) is an essential step for analyzing functional data \citep{ramsay1991some, ramsay2005functional}. However, the smoothing step has been neglected by most of the existing functional data analysis (FDA) methods, which integrate functional representations in the analysis models. For examples, functional data and effects are represented by basis functions in functional linear regression models \citep{cardot1999functional, cardot2003spline, hall2007methodology, Zhu200901, zhu2011robust}, functional additive models \citep{Fabian2015, fan2015}, functional principle components analysis \citep{crainiceanu2010bayesian, zhu2014structured}, and nonparametric functional regression models \citep{ferraty2006nonparametric, gromenko2013}; and represented by Gaussian processes (GP) in Bayesian nonparametric models \citep{gibbs1998bayesian, shi2007gaussian, banerjee2008gaussian, kaufman2010bayesian}.

On the other hand, most of the existing smoothing methods process one functional observation per time, such as cubic smoothing splines (CSS) and kernel smoothing \citep{green1993nonparametric, ramsay2005functional}. Consequently, when multiple functional observations are sampled from the same distribution, these individually smoothing methods lead to less accurate results, by ignoring the shared mean-covariance functions. Alternatively, \citet{yang2016} proposed a Bayesian hierarchical model (BHM) with Gaussian-Wishart processes for simultaneously and nonparametrically smoothing multiple functional observations and estimating mean-covariance functions, which is shown to be comparable with the frequentist method  --- Principle Analysis by Conditional Expectation (PACE) proposed by \citet{yao2005functional}.

BHM assumes a general measurement error model for the observed functional data $\{Y_i(t);\, t \in \mathcal{T}, \, i = 1, \cdots, n\}$,
\begin{equation}
\label{bhm_mod}
Y_i(t) = Z_i(t) + \epsilon_i(t); \; Z_i(\cdot) \sim GP(\mu_Z(\cdot), \Sigma_Z(\cdot, \cdot)),\; \epsilon_i(\cdot) \sim N(0, \sigma^2_{\epsilon});
\end{equation}
$$\mu_Z(\cdot)|\Sigma_Z(\cdot, \cdot) \sim GP\left(\mu_0(\cdot), \frac{1}{c}\Sigma_Z(\cdot, \cdot)\right), \; \Sigma_Z(\cdot, \cdot) \sim IWP(\delta, \sigma^2_sA(\cdot, \cdot)), \; \sigma^2_{\epsilon} \sim IG(a_{\epsilon}, b_{\epsilon}); $$
$$\sigma^2_s \sim IG(a_s, b_s); \; $$
where $\{Z_i(t); i = 1, \cdots, n\}$ denotes the underlying true functional data following the same GP distribution with mean function $\mu_Z(\cdot)$ and covariance function $\Sigma_Z(\cdot, \cdot)$, $IWP$ denotes the Inverse-Wishart process (IWP) prior \citep{dawid1981some} for the covariance function, $IG$ denotes the Inverse-Gamma prior, and $(\mu_0(\cdot), c, \delta, A(\cdot, \cdot), a_{\epsilon}, b_{\epsilon}, a_s, b_s)$ are hyper-prior parameters to be determined. The IWP prior on $\Sigma_Z(\cdot, \cdot)$ models the covariance function nonparametrically and hence allows the method for analyzing both stationary and nonstationry functional data with unknown covariance structures.

However, the BHM suffers serious computational burden and instability when functional data are observed on high-dimensional or random grids. To address the computational issue of Bayesian GP regression models for high-dimensional functional data, the existing reduce-rank methods focus on kriging with partial data \citep{cressie2008fixed, banerjee2008gaussian}, implementing direct low-rank approximations for the covariance matrix \citep{Rasmussen2006, Quinonero2007, shi2011, Banerjee2013}, and using predictive processes \citep{sang2012full, Finley2015}. Although these reduce-rank methods successfully apply to the standard GP regression models \citep{shi2007gaussian, banerjee2008gaussian, kaufman2010bayesian} that only model group-level GPs with parametric covariance functions, they greatly increase the complexity in BHM for handling multiple GPs (one per functional observation, one for the mean prior) and an IWP (prior for the covariance function).

In this paper, we propose a novel \textbf{B}ayesian framework with \textbf{A}pproximations by \textbf{B}asis \textbf{F}unctions for the original BHM method, referred to as \textbf{BABF}, which is computationally efficient and stable for analyzing high-dimensional functional data. Basically, we approximate the underlying true functional data $\{Z_i(t); i = 1, \cdots, n\}$ with basis functions, and derive an induced Bayesian hierarchical model on the basis-function coefficients from the original assumptions of BHM (\ref{bhm_mod}). Then we can conduct posterior inference for functional signals $\{Z_i(t); i = 1, \cdots, n\}$ and mean-covariance functions $(\mu_Z(\cdot), \Sigma_Z(\cdot, \cdot))$, by Markov chain Monte Carlo (MCMC) under the induced model of basis-function coefficients, i.e., by MCMC in the basis-function space with a reduced rank. As a result, our BABF method not only improves the computational scalability over the original BHM, but also inherits the advantage of modeling the functional data and mean-covariance functions in a flexible nonparametric manner. In addition, because of basis function approximations, BABF can naturally handle functional data observed on random or uncommon grids.

Thus, our basis function approximation approach has two-fold advantages: (i) Compared to the alternative reduce-rank approaches, it is easier to apply to Bayesian hierarchical GP methods that model individual levels of GPs (e.g., BHM). (ii) It induces a nonparametric Bayesian model with a Gaussian-Wishart prior for the basis-function coefficients, which is different from modeling the basis-function coefficients as independent variables as in the standard functional linear regression models \citep{cardot1999functional, cardot2003spline, hall2007methodology, Zhu200901, zhu2011robust} and functional additive models \citep{Fabian2015, fan2015}, and also different from directly modeling the basis-function coefficients in semiparametric forms as in \citet{baladandayuthapani2008bayesian}.

By simulation studies with both stationary and nonstationary functional data, we demonstrate that BABF produces accurate smoothing results and mean-covariance function estimates. Specifically, when functional data are observed on low-dimensional common grids, BABF generates similar results as the original BHM. When functional data are observed on high-dimensional or random grids, the original BHM fails because of computational issues, while BABF efficiently produces smoothed signal estimates with smaller root mean square errors (RMSEs) than the alternative methods (CSS, PACE).

Furthermore, by a real study with the sleeping energy expenditure (SEE) measurements of 106 children and adolescents (44 obese cases, 62 controls) over 405 time points \citep{LeeFPCAee}, we show that BABF captures better periodic patterns of the measurements, producing more reasonable estimates for the functional signals and  mean-covariance functions. Moreover, compared to the raw data and smoothed data by CSS and PACE, the smoothed data by BABF leads to better classification results for the SEE data.

This paper is organized as follows: We describe the BABF method in Section \ref{method}, present simulation and real studies in Sections \ref{simu} and \ref{app} respectively, and then conclude with a discussion in Section \ref{dis}.

\section{BABF method}
\label{method}

Because the original BHM method \citep{yang2016} conducts MCMC on the pooled observation grid for handling uncommon grids, it has computational complexity $O(np^3m)$ with $n$ samples, $p$ pooled-grid points, and $m$ MCMC iterations. To resolve the computational bottleneck issue for smoothing functional data with large pooled-grid dimension $p$ by BHM, we propose our BABF method by approximating functional data with basis functions under the same model assumptions as in BHM (\ref{bhm_mod}).

\subsection{Model description}
\label{babf}

First, we approximate the GP evaluations $\{Z_i(\bm{\tau})\}$ by a system of basis functions (e.g., cubic B-splines),  with a working grid based on data density, $\bm{\tau} = (\tau_1, \tau_2, \cdots, \tau_L)^T \subset \mathcal{T}$, $L<<p$. Let $B(\cdot) = [b_1(\cdot), b_2(\cdot), \cdots, b_K(\cdot)]$ denote $K$ selected basis functions with coefficients $\bm{\zeta_i} = (\zeta_{i1}, \zeta_{i2}, \cdots, \zeta_{iK})^T$, then
\begin{equation}
\label{splineZ}
Z_i(\bm{\tau}) = \sum_{k = 1}^K \zeta_{ik} b_k(\bm{\tau}) = \bm{B(\tau)} \bm{\zeta_i}. 
\end{equation}
Assuming $K=L$, we can write $\bm{\zeta_i} = \bm{B(\tau)}^{-1} Z_i(\bm{\tau})$ as a linear transformation of $Z_i(\bm{\tau})$. Note that even if $\bm{B(\tau)}$ is singular or non-square, $\bm{\zeta_i}$ can still be written as a linear transformation of $Z_i(\bm{\tau})$ with the generalized inverse \citep{james1978generalised} of $\bm{B(\tau)}$.
Consequently, the true signals $\{Z_i(\bm{t_i})\}$ can be approximated by
$\{\bm{B(t_i)} \bm{\zeta_i}\}$ with given $\{\bm{\zeta_i}\}$. 

Second, we derive the induced Bayesian hierarchical model for the basis-function coefficients $\{\bm{\zeta_i}\}$.  Because $\bm{\zeta_i}$ is a linear transformation of $Z_i(\bm{\tau})$ that follows a multivariate normal distribution $MN (\mu_Z(\bm{\tau}), \Sigma_Z(\bm{\tau}, \bm{\tau}))$ under the assumptions in (\ref{bhm_mod}), the induced model for $\bm{\zeta_i}$ is
\begin{equation}
\label{zeta}
\bm{\zeta_i} \sim MN(\bm{\mu_{\zeta}},\; \bm{\Sigma_{\zeta}}); \; \bm{\mu_{\zeta}} = \bm{B(\tau)}^{-1} \mu_Z(\bm{\tau}); \; \bm{\Sigma_{\zeta}} = \bm{B(\tau)}^{-1} \Sigma_Z(\bm{\tau}, \bm{\tau}) \bm{B(\tau)}^{-T}.
\end{equation} 
Further, from the assumed priors of $(\mu_Z(\cdot), \Sigma_Z(\cdot, \cdot))$ in (\ref{bhm_mod}), the following priors of $(\bm{\mu_{\zeta}}, \bm{\Sigma_{\zeta}})$ are also induced: 
\begin{eqnarray}
\bm{\mu_{\zeta}}| \bm{\Sigma_{\zeta}}  &\sim& MN\left(\bm{B(\tau)}^{-1} \mu_0(\bm{\tau}),\; c\bm{\Sigma_{\zeta}} \right); \label{mu_zeta_prior}\\
\bm{\Sigma_{\zeta}} &\sim& IW(\delta, \; \bm{B(\tau)}^{-1} \Psi(\bm{\tau}, \bm{\tau}) \bm{B(\tau)}^{-T}). \label{S_zeta_prior}
\end{eqnarray}

Last, we conduct MCMC by a Gibbs-Sampler \citep{geman1984stochastic} with computation complexity $O(nK^3m)$ under the above induced model of the basis-function coefficients. Details of the MCMC procedure are provided in Section \ref{mcmc}. We take the corresponding averages of the posterior MCMC samples as our Bayesian estimates, whose uncertainties can easily be quantified by the MCMC credible intervals.

\subsection{Hyper-prior selection}
\label{hyper_priors}

For setting hyper-priors, we use the same data-driven strategy as used by the original BHM method \citep{yang2016}. Specifically, we set $\mu_0(\cdot)$ as the smoothed sample mean, and $c=1$, $\delta=5$ for uninformative priors of the mean-covariance functions.
We set $A(\cdot, \cdot)$ as a Mat{\'e}rn covariance function \citep{matern1960spatial} for stationary data, or as a smooth covariance estimate for nonstationary data (e.g., PACE estimate, smoothed empirical estimate). 
A heuristic Bayesian approach is used for setting the values of $(a_{\epsilon}, b_{\epsilon}, a_s, b_s)$, by matching hyper-prior moments with the empirical estimates. 

\subsection{Basis-function selection}
\label{basis-func-selection}
The key feature of the BABF method is conducting MCMC with the induced model of the basis-function coefficients. BABF inherits the advantage of nonparametrically smoothing without the necessity of tuning smoothing parameters, where the amount of smoothness in the posterior estimates is determined by the data and the IWP prior of the covariance function. Therefore, the induced model of the basis-function coefficients makes BABF robust with respect to the selected basis functions and working grid. 

Moreover, the appropriately selected basis functions and working grid will help improve the performance of BABF. The general strategies of selecting basis functions for interpolating over the working grid apply here, where the basis-function type depends on the data, e.g., Fourier series for periodic data, B-splines for GP data, and wavelets for signal data. Using B-splines as an example, the optimal knot sequence for best interpolation at the working grid $\tau$ can be obtained using the method developed by \citet{gaffney1976optimal, micchelli1976optimal, de1977computational}, and implemented by the Matlab function \verb|optknt|.  The working grid $\tau$ can be chosen to represent data densities over the domain, e.g., given by the $\left(\frac{1}{L+1}, \cdots, \frac{L}{L+1}\right)$ percentiles of the pooled observation grid, or the equally-spaced grid for evenly distributed data. As for the dimension $L$ of the working grid, one may try a few values with a small testing data set, and then select one with the smallest RMSE of the signal estimates.

\subsection{Posterior inference}
\label{mcmc}
For the original BHM (\ref{bhm_mod}), the joint posterior distribution of $(\bm{Z}, \mu_Z, \Sigma_Z, \sigma_{\epsilon}^2, \sigma_s^2)$ is
\begin{eqnarray}
&f(\bm{Z}, \mu_Z, \Sigma_Z, \sigma_{\epsilon}^2, \sigma_s^2 | \bm{Y}) \propto f(\bm{Y} | \bm{Z}, \sigma_{\epsilon}^2) f(\bm{Z} | \mu_Z, \Sigma_Z) f(\mu_Z | \Sigma_Z)f(\Sigma_Z | \sigma_s^2)f(\sigma_{\epsilon}^2) f(\sigma_s^2),& \label{joint_post1} \\
&\bm{Z} = \{Z_1(\bm{t_i}), \cdots, Z_n(\bm{t_n})\}, \; \bm{Y} =\{Y_1(\bm{t_i}), \cdots, Y_n(\bm{t_n})\}. & \nonumber
\end{eqnarray}
Equivalently, because of $\bm{\zeta_i} = \bm{B(\tau)}^{-1} Z_i(\bm{\tau})$, the joint posterior distribution of $(\bm{\zeta}, \bm{\mu_{\zeta}}, \bm{\Sigma_{\zeta}}, \sigma_{\epsilon}^2, \sigma_s^2)$  is
\begin{eqnarray} 
f(\bm{\zeta}, \bm{\mu_{\zeta}}, \bm{\Sigma_{\zeta}}, \sigma_{\epsilon}^2, \sigma_s^2 | \bm{Y}) \propto f(\bm{Y} | \bm{\zeta}, \sigma_{\epsilon}^2) f(\bm{\zeta} |\bm{\mu_{\zeta}}, \bm{\Sigma_{\zeta}} ) f(\bm{\mu_{\zeta}}| \bm{\Sigma_{\zeta})} f(\bm{\Sigma_{\zeta}} | \sigma_s^2)f(\sigma_{\epsilon}^2) f(\sigma_s^2), \label{joint_post2} \\
\bm{\zeta}=\{\bm{\zeta_1}, \cdots, \bm{\zeta_n}\},\; \bm{\mu_{\zeta}} = \bm{B(\tau)}^{-1} \mu_Z(\bm{\tau}), \;\bm{\Sigma_{\zeta}} = \bm{B(\tau)}^{-1} \Sigma_Z(\bm{\tau}, \bm{\tau}) \bm{B(\tau)}^{-T}.  \nonumber
\end{eqnarray} 

\subsubsection{Full conditional distribution of $\bm{\zeta_i}$}

From (\ref{joint_post2}), we can see that 
$$f(\bm{\zeta} | \bm{Y}, \bm{\mu_{\zeta}}, \bm{\Sigma_{\zeta}}) \propto f(\bm{Y} | \bm{\zeta}, \sigma_{\epsilon}^2) f(\bm{\zeta} |\bm{\mu_{\zeta}}, \bm{\Sigma_{\zeta}} ).$$
Then  the full conditional posterior distribution of $\bm{\zeta_i}$ is derived as
\begin{eqnarray}
&\bm{\zeta_i} | (Y_i(\bm{t_i}), \bm{\mu_{\zeta}}, \bm{\Sigma_{\zeta}} )
\sim MN\left[  \bm{m_{\zeta_i | Y_i}}, \; \bm{V_{\zeta_i| Y_i} } \right];& \label{zetai:cond} \\
&\bm{V_{\zeta_i| Y_i} } = \left( \frac{\bm{B(t_i)}^T \bm{B(t_i)}}{\sigma_{\epsilon}^2} + \bm{\Sigma_{\zeta}}^{-1}  \right)^{-1},\;
 \bm{m_{\zeta_i | Y_i}} = \bm{V_{\zeta_i| Y_i} } \left( \frac{\bm{B(t_i)}^T Y_i(\bm{t_i})}{\sigma_{\epsilon}^2} + \bm{\Sigma_{\zeta}}^{-1} \bm{\mu_{\zeta}} \right).& \nonumber
\end{eqnarray}

\subsubsection{Full conditional distribution for $\bm{\mu_{\zeta}}$, $\bm{\Sigma_{\zeta}}$}
Conditioning on $\{\bm{\zeta}_i\}$, the posterior distribution of $(\bm{\mu_{\zeta}}$, $\bm{\Sigma_{\zeta}})$ is
$$ f(\bm{\mu_{\zeta}}, \bm{\Sigma_{\zeta}} | \bm{\zeta_1}, \dots, \bm{\zeta_n}) \propto 
 \prod_{i=1}^nf(\bm{\zeta_i} | \bm{\mu_{\zeta}}, \bm{\Sigma_{\zeta}}) f(\bm{\mu_{\zeta}} |\bm{\Sigma_{\zeta}})f(\bm{\Sigma_{\zeta}}), $$ 
where $f(\bm{\mu_{\zeta}}|\bm{\Sigma_{\zeta}})$ and $f(\bm{\Sigma_{\zeta}})$ are given by (\ref{mu_zeta_prior}), (\ref{S_zeta_prior}).
Therefore, 
\begin{eqnarray}
&\bm{\mu_{\zeta}} | (\bm{\zeta_1}, \dots, \bm{\zeta_n}, \bm{\Sigma_{\zeta}}) 
\sim MN\left( \frac{1}{n+c}\left(\sum_{i=1}^n \bm{\zeta_i} + c \bm{B(\tau)}^{-1} \mu_0(\bm{\tau})\right),\; \frac{1}{n+c} \bm{\Sigma_{\zeta}}  \right); & \label{mu_zeta_post} \\
&\bm{\Sigma_{\zeta}} | (\bm{\zeta_1}, \dots, \bm{\zeta_n}, \bm{\mu_{\zeta}}) \sim
IW(\widetilde{\delta_{\zeta}},\; \widetilde{\bm{\Psi_{\zeta}}}),\;& \label{S_zeta_post} \\
&\widetilde{\delta_{\zeta}} = n+1+\delta,\; \widetilde{\bm{\Psi_{\zeta}}} =  \sum_{i=1}^n (\bm{\zeta_i} - \bm{\mu_{\zeta}})(\bm{\zeta_i} - \bm{\mu_{\zeta}})^T + &  \nonumber\\
&c(\bm{\mu_{\zeta}} - \bm{B(\tau)}^{-1}\mu_0(\bm{\tau}))(\bm{\mu_{\zeta}} - \bm{B(\tau)}^{-1}\mu_0(\bm{\tau}))^T + \bm{B(\tau)}^{-1}\Psi(\bm{\tau}, \bm{\tau})\bm{B(\tau)}^{-T}.& \nonumber 
\end{eqnarray}

\subsubsection{MCMC procedure}
\label{mcmc}
We design the following Gibbs-Sampler algorithm for MCMC, which ensures computational convenience and posterior convergence. 

\noindent Step 0: Set hyper-priors (Section \ref{hyper_priors}) and initial parameter values. Initial values for $(\mu_Z(\bm{\tau}), \Sigma_Z(\bm{\tau}, \bm{\tau}), \sigma^2_{\epsilon})$ can be set as empirical estimates, inducing the initial values for $(\bm{\mu_{\zeta}}, \bm{\Sigma_{\zeta}})$ by (\ref{zeta}).

\noindent Step 1: Conditioning on observed data $\bm{Y}$ and the current values of $(\bm{\mu_{\zeta}}, \bm{\Sigma_{\zeta}}, \sigma_{\epsilon}^2)$, sample $\{\bm{\zeta_i}\}$ from (\ref{zetai:cond}).

\noindent Step 2: Conditioning on the current values of $\bm{\zeta}$, update $\bm{\mu_{\zeta}}$ and $\bm{\Sigma_{\zeta}}$ respectively from (\ref{mu_zeta_post}) and (\ref{S_zeta_post}).

\noindent Step 3: Given the current values of $(\{\bm{\zeta_i}\}, \bm{\mu_{\zeta}}, \bm{\Sigma_{\zeta}})$, approximate $\{Z_i(\bm{t_i})$, $\mu_Z(\bm{t_i})$, $\Sigma_Z(\bm{t_i}, \bm{t_i})$, $\Sigma(\bm{\tau}, \bm{t_i}), \Sigma_Z(\bm{t_i}, \bm{\tau}), \Sigma_Z(\bm{\tau}, \bm{\tau})\}$ by 
$$Z_i(\bm{t_i}) = \bm{B(t_i)} \bm{\zeta_i}, \; \mu_Z(\bm{t_i}) = \bm{B(t_i)\mu_{\zeta}}, \;  \Sigma(\bm{t_i}, \bm{t_i}) =  \bm{B(t_i)}\bm{\Sigma_{\zeta}} \bm{B(t_i)}^T, \; $$
$$\Sigma(\bm{\tau}, \bm{t_i})^T = \Sigma(\bm{t_i}, \bm{\tau}) =  \bm{B(t_i)}\bm{\Sigma_{\zeta}} \bm{B(\tau)}^T, \; 
\Sigma(\bm{\tau}, \bm{\tau}) =  \bm{B(\tau)}\bm{\Sigma_{\zeta}} \bm{B(\tau)}^T .$$

\noindent Step 4: Conditioning on $\bm{Z}$ and $\bm{Y}$, update $\sigma_{\epsilon}^2$ by
$$IG\left( a_{\epsilon} + \frac{1}{2} \sum_{i=1}^n p_i ,  \;
b_{\epsilon} + \frac{1}{2} \sum_{i=1}^n  (Y_i(\bm{t_i}) - Z_i(\bm{t_i}))^T(Y_i(\bm{t_i}) - Z_i(\bm{t_i}))   \right),$$
which is derived from 
$$f(\sigma_{\epsilon}^2 | Y_1(\bm{t_1}), Z_1(\bm{t_1}),  \cdots,  Y_n(\bm{t_n}), Z_n(\bm{t_n})) \propto \prod_{i=1}^n f(Y_i(\bm{t_i}) | Z_i(\bm{t_i}), \sigma_{\epsilon}^2) f(\sigma_{\epsilon}^2).$$

\noindent Step 5: Given the current value of $\bm{\Sigma_{\tau}} = \Sigma_Z(\bm{\tau}, \bm{\tau})$, update $\sigma_s^2$   by 
$$ \sigma_s^2 | \bm{\Sigma_{\tau}} \sim G\left(a_s + \frac{(\delta+K-1)K}{2}, \; b_s + \frac{1}{2} trace(\bm{A(\tau, \tau)\Sigma_{\tau}}^{-1})\right), $$
which is derived from 
$$f(\sigma_s^2 | \bm{\Sigma_{\tau}}) \propto f(\bm{\Sigma_{\tau}} | \sigma_s^2) f(\sigma_s^2).$$

In general, the posterior samples will pass the convergence diagnosis by potential scale reduction factor (PSRF) \citep{gelman1992inference}, with a fairly large number of MCMC iterations (e.g., 12,000 in our numerical studies).

\section{Simulation studies}
\label{simu}

In the following simulation studies, we compared the BABF method with CSS \citep{green1993nonparametric}, PACE \citep{Yao2005}, Bayesian functional principle component analysis (BFPCA) \citep{crainiceanu2010bayesian}, standard Bayesian GP regression (BGP) \citep{gibbs1998bayesian}, and the original BHM method \citep{yang2016}. We considered scenarios with stationary and nonstationary functional data, common and random observation grids, Gaussian and non-Gaussian data. Because both BFPCA and BGP are developed for the scenario with common grids; BHM has computational issues with high-dimensional pooled-grid (the case with random grids); and BHM is known to be comparable with PACE \citep{yang2016}. We compared all methods in the scenario with common grids, but only compared BABF with CSS and PACE in the scenario with random grids. 

Because simulation data were evenly distributed over the domain, we selected an equally spaced working grid with length $20$ for BABF. CSS was applied to each functional observation independently with the smoothing parameter selected by general cross-validation (GCV). For BFPCA, we used the covariance estimate by PACE, and selected the number of principle functions subject to capture $99.99\%$ data variance.   For BGP, we assumed the Mat\'ern model for the covariance function with stationary data, while fixing the covariance at the PACE estimate with nonstationary data. 
All MCMC samples consisted of $2,000$ burn-ins and $10,000$ posterior samples, and passed the convergence diagnoses by PSRF \citep{gelman1992inference}.

\subsection{Studies with common grids}
\label{common_grid}

We generated $30$ stationary functional curves (true signals) on the common equally-spaced-grid with length $40$, over $\mathcal{T} = (0, \pi/2)$, from  
\begin{equation}
\label{st:gp}
GP( \mu(t) = 3\sin(4t), \Sigma(s, t) = 5Matern_{cor}(|s - t|;\rho = 0.5,  \nu = 3.5)),
\end{equation}
denoted by $\mathbf{Z}$. Specifically, 
$$
 Matern_{cor}(d;\rho,\nu) =  \frac{1}{\Gamma(\nu) 2^{\nu - 1}}
\left(\sqrt{2\nu}\frac{d}{\rho}  \right)^{\nu}  K_{\nu}\left( \sqrt{2\nu}\frac{d}{\rho}\right), d\ge 0,\; \rho>0, \; \nu>0,
$$
where $\rho$ is the scale parameter, $\nu$ is the order of smoothness, $\Gamma(\cdot)$ is the gamma function, and $K_\nu(\cdot)$ is the modified Bessel function of the second kind. The noise terms $\{\epsilon_{ij}\}$ were generated from $N(0,  \sigma_{\epsilon} = \sqrt{5} / 2)$, such that the signal to noise ratio (SNR) was 2 (resulting relatively high volume of noise in the simulated data). The observed noisy functional data curves  were given by
$\mathbf{Y} = \mathbf{Z}+ \bm{\epsilon}$.

Similarly, we generated $30$ nonstationary functional curves on the same equally-spaced-grid with length $40$, from a nonstationary GP $\widetilde{X}(t) = h(t)X(s(t))$ (i.e., a nonlinear transformation of a stationary GP $X(\cdot)$), where $X(\cdot)$ denotes the GP (\ref{st:gp}), $h(t) = t + 1/2$, $s(t) = t^{2/3}$. Noisy observation data were obtained by adding noises from $N(0,  \sigma_{\epsilon} = \sqrt{5} / 2)$ to the generated nonstationary GP data (true signals).

We repeated the simulations 100 times, and calculated the RMSEs of the estimates of signals $\{Z_i(\bm{t})\}$, mean function $\mu_Z(\bm{t})$, covariance surface $\Sigma_Z(\bm{t}, \bm{t})$, and residual variance $\sigma^2_{\epsilon}$ ($\bm{t}$ denotes the common observation grid). The average RMSEs (with standard deviations among these 100 simulations) for stationary and nonstationary data were shown in Table \ref{mse_cgrid}, where the CSS estimates of $(\mu_Z, \Sigma_Z)$ were sample estimates with pre-smoothed signals by CSS, and average RMSEs were omitted if the parameters were not directly estimated by the corresponding methods, e.g., $(\mu_Z, \Sigma_Z, \sigma^2_{\epsilon})$ by BFPCA, $\sigma^2_{\epsilon}$ by CSS.  

Table \ref{mse_cgrid} shows that BGP produces the best estimates for the signals and residual variance (with the lowest RMSEs), while BHM and BABF gives the second best estimates for the signals and residual variance, as well as the best estimates for the mean-covariance functions. With nonstationary data of common grids, BGP and PACE produce the best covariance estimates, while BABF produces closely accurate covariance estimates, as well as the best estimates for the signals, mean function, and residual variance. Because of stable computations with nonstationary data, our BABF method produces  better estimates than BHM. In addition,  the CSS and BFPCA methods produce the least accurate estimates (with the highest RMSEs) for both stationary and nonstationary data, which demonstrates the advantage of simultaneously smoothing and estimating functional data as in BGP, BHM, and BABF.

\begin{table}[ht]
\caption{Simulation results with common grids: average RMSEs and corresponding standard errors (in parentheses) of $\{Z_i(\bm{t})\}$, $\mu(\bm{t})$, $\Sigma_Z(\bm{t}, \bm{t})$, and $\sigma^2_{\epsilon}$ produced by CSS, PACE, BFPCA, BGP, BHM, and BABF. Average RMSEs are omitted if the corresponding parameters are not directly estimated. Two best results are bold for each parameter.}
\begin{center} {
\begin{tabular}{lllllllll}
\hline
 &  CSS & PACE & BFPCA & BGP & BHM & BABF\\
\hline
Stationary & &&&&&& \\
$\{Z_i(\bm{t})\}$  & 0.4808 & 0.4553 & 0.5657 & \textbf{0.4020} & \textbf{0.4067}  & 0.4073 \\
          &(0.0213) &(0.0268)& (0.0550) & (0.0219) & (0.0207) & (0.0204)\\

 $\mu(\bm{t})$ & 0.4757 & 0.4194 & -  &  0.3982  &\textbf{0.3961} &\textbf{0.3961} \\
         & (0.1347) &(0.1593) & - &(0.1527)  &(0.1538) & (0.1535) \\

 $\Sigma(\bm{t}, \bm{t})$ & 1.0017 & 1.0375 & -  &  1.0988 &\textbf{0.9601} &\textbf{0.9590} \\
              & (0.3079) &(0.2850) & - &(0.4934)  &(0.2902) & (0.2913) \\


$\sigma^2_{\epsilon}$ & - & 0.0764 & - & \textbf{0.0460}  &0.0491     &\textbf{0.0483}    \\
                      &-  &(0.0516)  & - &(0.0327)      & (0.0357)      & (0.0352)  \\
Nonstationary  & &&&&&& \\
$\{Z_i(\bm{t})\}$  & 1.0271 & 0.5185 & 0.6314 & \textbf{0.5183} & 0.5759 &\textbf{0.5133} \\
          &(0.00463) &(0.0255)& (0.0632) & (0.0265) & (0.0227) & (0.0227)\\

$\mu(\bm{t})$ & 0.9446  & 0.5782 & -  &  \textbf{0.5387}  &0.5530 &\textbf{0.5356} \\
         & (0.1509) &(0.2095) & - &(0.2090)  &(0.2038) & (0.2094) \\

$\Sigma(\bm{t}, \bm{t})$ & 1.9635 &\textbf{1.9751}& - &\textbf{1.9733} & 2.0296 &1.9768 \\
              & (0.8386) &(0.8160) & - &(0.6831)  &(0.6891) & (0.7835) \\

$\sigma^2_{\epsilon}$ & - & \textbf{0.0810} & - & 0.1472 & 0.2432  &\textbf{0.0692}    \\
            &-  &(0.0541)  & - &(0.0879)     & (0.0644)  & (0.0492)  \\
\hline
\end{tabular}}
\end{center}
\label{mse_cgrid}
\end{table}

Figure \ref{fig:1} (a, b, c, d) shows that all three Bayesian methods produce similarly accurate estimates for the functional signals and mean function of common grids. With nonstationary data, our BABF method produces the best signal estimates (Figure \ref{fig:1}(b)).
As for the functional covariance estimates (Supplementary Figure 1), the parametric estimate by BGP is a Mat\'ern function because of the assumed true Mat\'ern  covariance model, but with underestimated diagonal variances. Practically, a wrong covariance model is usually assumed in BGP, which is likely to produce estimates with large errors and wrong structures. In contrast, the nonparametric methods such as BHM and BABF are more flexible and applicable for estimating the covariance function of real data.

In addition, we examined the coverage probabilities of the 95\% pointwise credible intervals (CI) generated by BGP, BHM, and BABF, for the functional signals and mean-covariance functions (Supplementary Table 1). For functional signals, BGP results the highest coverage probability  with stationary data ($0.9483$ vs.~$0.9217, 0.9208$), but the lowest coverage probability with nonstationary data ($0.8350$ vs.~$0.9450, 0.8742$). All methods have similar coverage probabilities for the functional mean ($\sim0.7$), where the relatively low coverage probabilities are due to the narrow 95\% confidence intervals. As for the covariance, the coverage probability by BGP is significantly lower than the ones by BHM and BABF for both stationary ($0.000$ vs.~$0.7869, 0.7869$) and nonstationary data ($0.3819$ vs.~$0.9913, 0.9938$), because BGP underestimates the diagonal variances.  

In summary, with common grids, Bayesian GP based regression methods (BGP, BHM, and BABF) produce better smoothing and estimation results, compared to estimating mean-covariance functions using the pre-smoothed functional data by CSS. Moreover, the results by BABF are at least similar to the ones by the original BHM, and better with nonstationary data.

\subsection{Studies with random grids}
\label{random_grid}

For this set of simulations, 
we generated $30$ true functional curves from the  
stationary and non-stationary GPs as in Section \ref{common_grid}, with observational grids (length $40$) that were randomly (uniformly) generated over $\mathcal{T} = (0, \pi/2)$. Raw functional data were then obtained by adding noises from $N(0, \sqrt{5}/2)$ to the true signals. We compared our BABF method (using an equally spaced working grid $\bm{\tau}_{1\times 20} \subset \mathcal{T}$) with CSS and PACE, by 100 simulations.

Table \ref{mse_rgrid} presents the average RMSEs for the estimates of the signals, residual variance, and mean-covariance functions (evaluated on the equally spaced grid over $\mathcal{T}$ with length $40$), along with standard errors in the parentheses. It is shown that our BABF method (with lowest RMSEs) performs consistently better than CSS and PACE for signal and mean estimates, with both stationary and nonstationary data of random grids.

\begin{table}[ht]
\caption{Simulation results with random grids: average RMSEs and corresponding standard errors (in parentheses) of $\{Z_i(\bm{t})\}$, $\mu(\bm{t})$, $\Sigma_Z(\bm{t}, \bm{t})$, and $\sigma^2_{\epsilon}$ by CSS, PACE, and BABF. Average RMSEs are omitted if the corresponding parameters are not directly estimated. Best results are bold for each parameter.}
\begin{center} {
\begin{tabular}{cccc||ccc}
\hline
 &  \multicolumn{3}{c}{Stationary}   & \multicolumn{3}{c}{Nonstationary} \\

\hline
  &   CSS & PACE & BABF & CSS & PACE & BABF\\
\hline
$\{Z_i(\bm{t})\}$  &  0.4839 & 1.4141  & \textbf{0.4079} &  1.0137 & 2.6300  & \textbf{0.6832}  \\
        &(0.0229) &  (0.1424)&  (0.0219) & (0.0511) &  (0.2876) &   (0.0576)\\

 $\mu(\bm{t})$ & 0.4229 & 0.4196   & \textbf{0.3690}  & 0.9905  &  0.6157 &  \textbf{0.5920} \\
         & (0.1471) & (0.1290) & (0.1302) & (0.1888)  & (0.2160)  & (0.2138)\\

 $\Sigma(\bm{t}, \bm{t})$ & 1.0445 &  1.4089 & \textbf{1.0054 } & \textbf{1.6403} & 2.4120  & 2.2090 \\
            & 0.4313  & (0.3502) & (0.3286) & (0.6086) &  (0.6497) &  (0.4506)   \\

$\sigma^2_{\epsilon}$ & - & 0.1900& \textbf{0.0509}   &  -  & 0.4007 &  \textbf{0.2209}\\
            & - & (0.1818) & (0.0387) & -  &  (0.2960) &  (0.1189) \\
\hline
\end{tabular}}
\end{center}
\label{mse_rgrid}
\end{table}

\begin{figure}[ht]
\begin{center}
\includegraphics[width=0.8\textwidth]{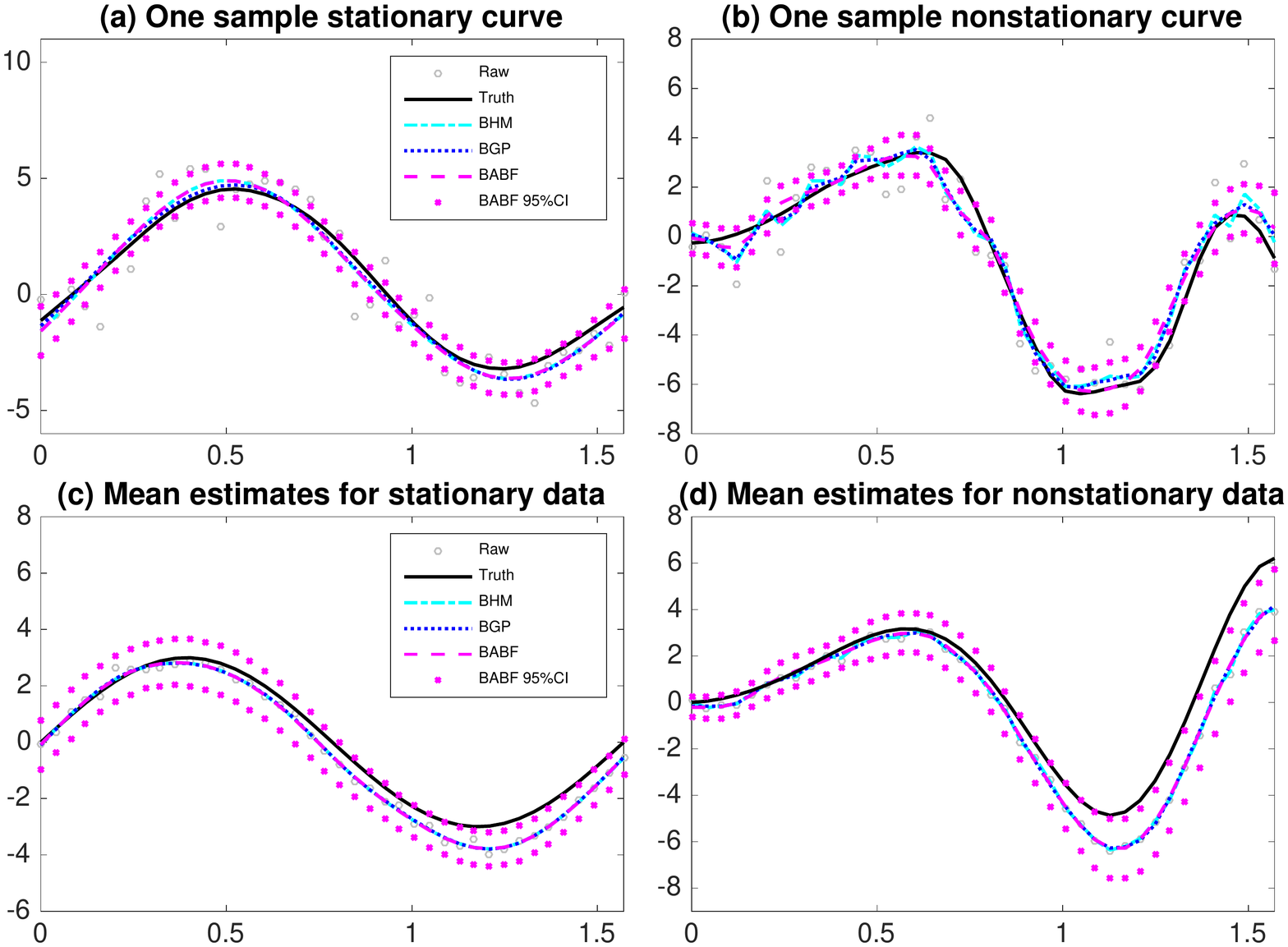}
\includegraphics[width=0.8\textwidth]{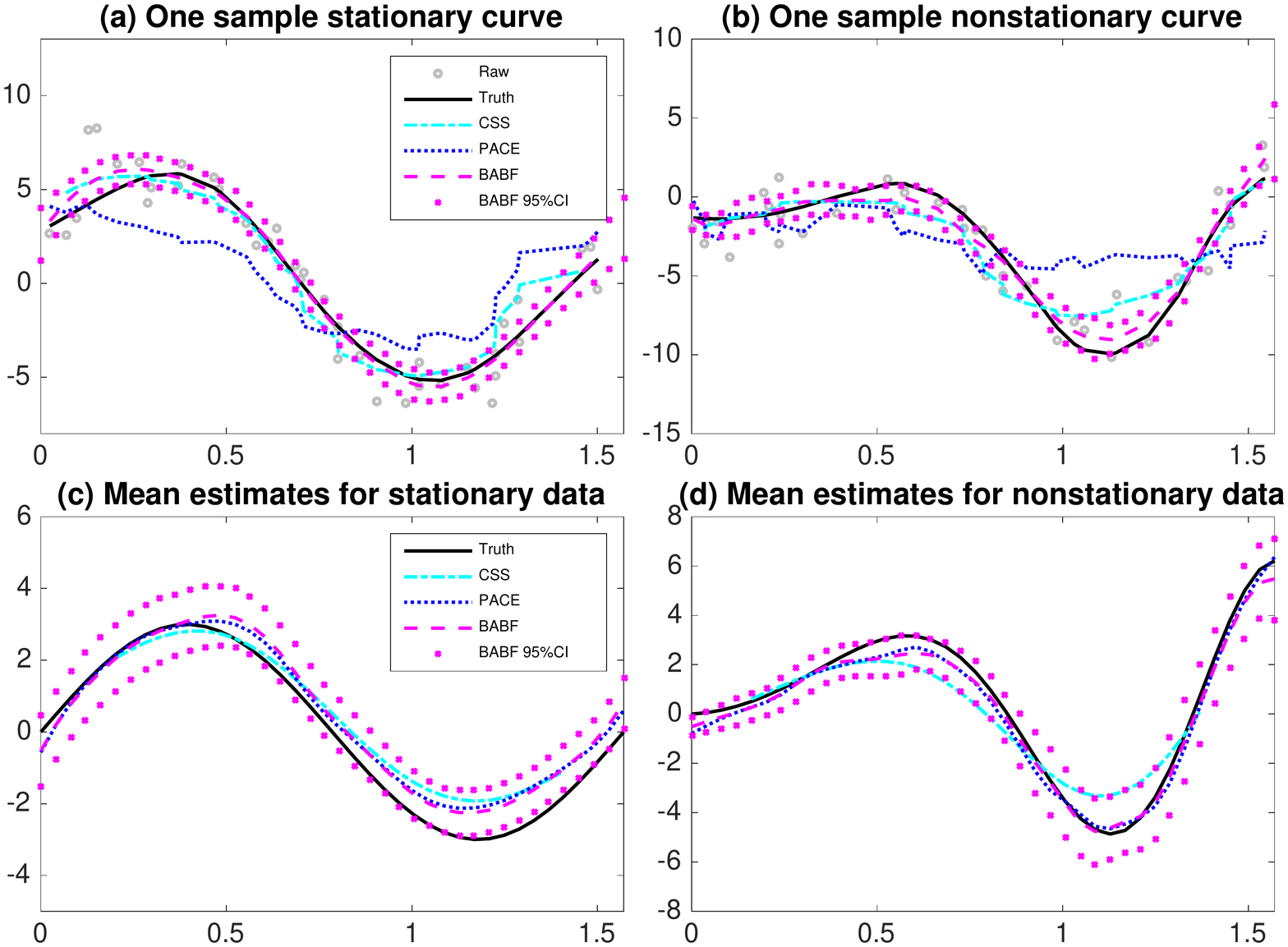}
\caption{Example smoothed functional data of common grids in (a, b), mean estimates of common grids in (c, d), example smoothed functional data of random grids in (e, f), and mean estimates of random grids in (g, h), along with 95\% pointwise CIs by BABF.}
\label{fig:1}
\end{center}
\end{figure}

Figure \ref{fig:1} (e, f) shows that BABF produces the best signal estimates in the scenario with random grids.
This is because CSS smoothed each functional curve independently; PACE only uses limited information per pooled-grid point; while BABF borrows strength across all observations through basis function approximations. For both stationary and nonstationary functional data, PACE and BABF give closely accurate mean estimates, while CSS gives the least accurate mean estimate (Figure \ref{fig:1} (g, h)). In addition, PACE produces the roughest covariance estimate (Supplementary Figure 2), for only using limited information on the pooled-grid points. The BABF coverage probability of the covariance is $0.9506$ for stationary data and $0.8550$ for nonstationary data, showing the good performance of our BABF method.

In summary, with random grids, our BABF method produces the best signal and mean estimates, compared to CSS and PACE. Although the sample covariance estimate using the pre-smoothed data generated by CSS has the lowest RMSE for nonstationary data, the analogous estimate using the more accurately smoothed data generated by BABF will have at least similar RMSE.

\subsection{Studies about robustness}
To test the robustness of our BABF method for handling non-Gaussian data, we further simulated stationary functional data from a non-Gaussian process, $0.2 (X(t)^2 - 1) + X(t)$, which is a modified Hermite polynomial transformation of the GP $X(t)$ in (\ref{st:gp}). We simulated functional data with $n=30$, random grids ($p=40$) over $\mathcal{T} = (0, \pi/2)$, and noises from $N(0, \sqrt{5}/2)$. Compared to CSS, our BABF method has RMSE $0.4278$ vs.~$0.7092$ for the signal estimates, $0.1271$ vs.~$0.4992$ for the functional mean estimate, and $0.4417$ vs.~$0.8886$ for the functional covariance estimate. These results demonstrate that our BABF method is robust for analyzing non-Gaussian functional data. In addition, we note that it is crucial to select a correct prior structure, $A(\cdot, \cdot)$ in (\ref{bhm_mod}), for the functional covariance. In general, we suggest using the Mat{\'e}rn model for stationary data and a smoothed covariance estimate by PACE for nonstationary data.

\subsection{Goodness-of-fit diagnostics}
We applied the method of goodness-of-fit \citep{yuan2012} using pivotal discrepancy measures (PDMs) on the residuals, $\epsilon_i(t) = Y_i(t) - Z_i(t)$, to examine the global goodness-of-fit of the Bayesian hierarchical model (\ref{bhm_mod}). As functions of the data and model parameters, the PDMs have the same invariant distribution when evaluated at the data-generating parameter value and parameter values drawn from the posterior distribution. Following the method proposed by \citet{yuan2012}, we constructed PDMs using standardized residuals from the posterior samples in MCMC. The PDM follows a chi-squared distribution under the null hypothesis that the residuals follow the $N(0, \sigma^2_{\epsilon})$ distribution (i.e., global goodness-of-fit for the Bayesian hierarchical model). In all simulation studies, the p-values of testing the null hypothesis of global goodness-of-fit for the Bayesian hierarchical model are greater than $0.25$, providing no evidence of lack-of-fit.

\section{Application on real data}
\label{app}

We analyzed a functional dataset
 from an obesity study with children and adolescents \citep{LeeFPCAee}, by the Children's Nutrition Research Center (CNRC) at Baylor College of Medicine.  This study estimated the energy expenditure (EE in unit kcal) of 106 children and adolescents (44 obese cases, 62 nonobese controls) during 24 hours with a series of scheduled physical activities and a sleeping period (12:00am-7:00am), by using the CNRC room respiration calorimeters \citep{moon1995closed}. We only analyzed the sleeping energy expenditure (SEE) data measured at 405 time points during the sleeping period. This real SEE data set provides a good example of high-dimensional common grids. The goal of this study was to discover different data patterns between obese cases and controls, providing insights about obesity diagnosis.

 We applied CSS, PACE, and our BABF method on this SEE functional data. Specifically, CSS was applied independently per sample with a smoothing parameter selected by GCV; PACE was applied with common grid $[1:405]$; and BABF was applied with the equally spaced working grid over $[1:405]$ with length $30$. Both PACE and BABF were applied separately for the functional data of obese and nonobese groups. Figure \ref{fig:2} (a, b) shows that CSS produces the roughest signal estimates, leading to the roughest mean-covariance estimates (Figures \ref{fig:2} (c, d); Supplementary Figures 3 and 4). Both PACE and BABF produce smoothed signal estimates and mean-covariance estimates. The mean estimate by BABF has better periodic patterns than the one by PACE (Figures 5 (c, d)), and the BABF estimates of the correlations between two apart time points are less than the PACE estimates (Supplementary Figure 4). 

\begin{figure}[ht]
\begin{center}
\includegraphics[width=\textwidth ]{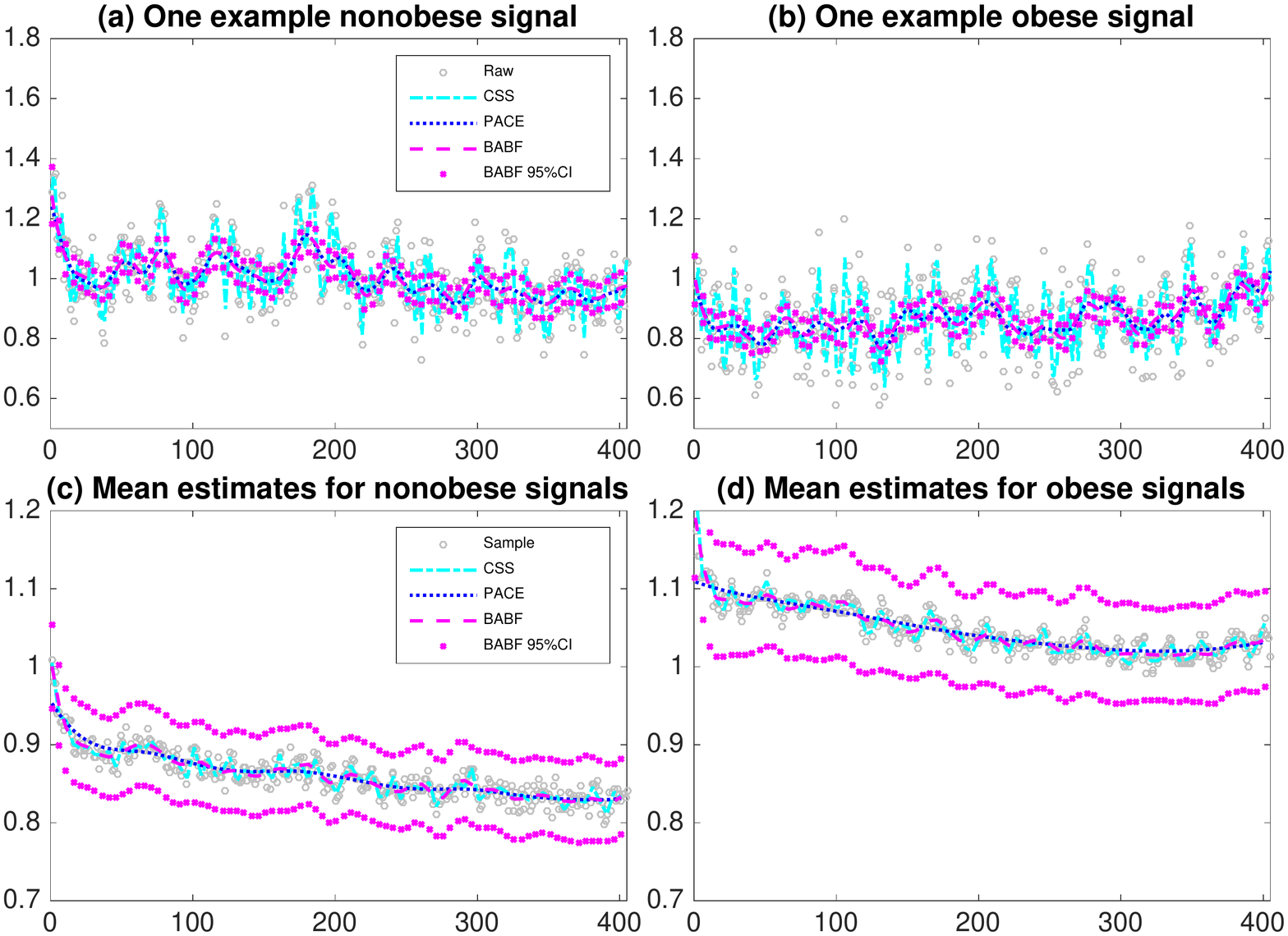}
\caption{ Example smoothed functional data in (a, b) and mean estimates in (c, d), along with 95\% pointwise CIs by BABF, for the real SEE data.}
\label{fig:2}
\end{center}
\end{figure}

Further, we applied the goodness-of-fit test \citep{yuan2012} to the residuals from the BABF method (one test per functional sample). Although the residual means are consistently close to $0$, the p-values for 52\% functional curves are less than $0.05/n$, suggesting evidences of lack-of-fit with Bonferroni correction \citep{bonferroni1936teoria} for multiple testing. This is because the residual variances of this real data are no longer the same across all observations. To address the issue of lack-of-fit for this SEE data, we need to assume sample-specific residual variances in the Bayesian hierarchical model (\ref{bhm_mod}), which is beyond the scope of this paper but will be part of our future research. 

Despite the lack-of-fit issue for this real data application by BABF, the smoothed data by BABF are improved over the raw data and the smoothed data by alternative methods for follow-up analyses. Using classification analysis as an example, we next illustrate the advantage of using the smoothed data by BABF for follow-up analyses. Considering the SEE data of obese and nonobese children as two classes, we used the leave-one-out cross-validation (LOOCV) approach to evaluate the classification results for using the raw data, and the smoothed data by CSS, PACE, and BABF. Basically, for each sample curve, we trained a SVM model \citep{cortes1995support} using the other sample curves, and then predicted if the test sample was an obese case. The error rate (the proportion of misclassification out of 106 samples) is $48.11\%$ for using the raw data, $40.57\%$ for using the smoothed data by CSS, and $36.79\%$ for using the smoothed data by PACE, and $33.02\%$ for using the smoothed data by BABF. The smoothed data by our BABF method lead to the smallest error rate. Thus, we believe using the smoothed data by BABF will be useful for follow-up analyses.


\section{Discussion}
\label{dis}

In this paper, we propose a computational efficient Bayesian method (BABF) for smoothing and estimating mean-covariance functions of high-dimensional functional data, improving upon the previous BHM method by \cite{yang2016}. Our BABF method projects the original functional data onto the space of selected basis functions with reduced rank, and then conducts posterior inference through MCMC of the basis-function coefficients. 
As a result, BABF method not only retains the same advantages as BHM, such as simultaneously smoothing and estimating mean-covariance functions in a nonparametric way, but also provides additional computational advantages of scalability, efficiency, and stability. A software for implementing the BHM and BABF methods is freely available at \url{https://github.com/yjingj/BFDA} \citep{yang2016bfda}.

With $n$ functional observations, a pooled observation grid of dimension $p$, and $m$ MCMC iterations, BABF reduces the computational complexity from $O(np^3m)$ to $O(nK^3m)$, and the memory usage from $O(p^2m)$ to $O(K^2m)$, by MCMC in the basis-function space with reduced rank $K<<p$.  
For examples, using a 3.2 GHz Intel Core i5 processor, BABF only costs about 3 minutes for $n=30$, $K=20$, and $m=12,000$, and about 9 minutes for $n=44$, $K=30$, and $m=12,000$. Although BABF (with $12,000$ MCMC iterations) costs about 4x longer than PACE, BABF provides complementary credible intervals to quantify the uncertainties of the posterior estimates, as well as basis function representations for the nonparametric estimates of functional signals and mean-covariance functions. Moreover, BABF produces more accurate results than PACE for functional data observed on random grids.

Both simulation and real studies demonstrate that BABF performs similarly as BHM and other Bayesian GP regression methods with functional data observed on low-dimensional common grids, and that BABF outperforms the alternative methods (e.g., CSS, PACE) with functional data observed on random grids or high-dimensional common grids. In addition, the real application shows that the classification analysis using the smoothed data by BABF produces the most accurate results. 

BABF assumes the same mean-covariance functions and residual variance for functional data, both of which are not true for most of the real data. Despite the model inadequacy, the smoothed data by BABF are still useful for follow-up analyses as shown in the real application of SEE data. To make the method more flexible for real data analysis, one might assume group-specific mean-covariance functions and sample-specific residual variances. This is beyond the scope of this paper and will be part of our future research.

In conclusion, BABF greatly improves the computational scalability and decreases the memory usage upon the original BHM method, while efficiently smoothing functional data and estimating mean-covariance functions in a nonparametric way. By implementing MCMC with the induced model of basis-function coefficients, our novel basis function approximation approach provides one solution for the computational bottleneck of general Bayesian GP regression methods, especially for analyzing high-dimensional functional data with Gaussian-Wishart processes.

\section*{Acknowledgments}
The authors would like to thank the Children's Nutrition Research Center at the Baylor College of Medicine for providing the metabolic SEE data (funded by National Institute of Diabetes and Digestive and Kidney Diseases Grant DK-74387 and the USDA/ARS under Cooperative Agreement 6250-51000-037). The authors would like to thank the writing lab of the School of Public Health at University of Michigan for helping proofread this manuscript. Jingjing Yang and Dennis D.~Cox were supported by the NIH grant PO1-CA-082710.

\bibliographystyle{biom}
\bibliography{BAsmooth}
\label{lastpage}
\newpage
}
\end{document}